\documentclass[conference]{IEEEtran}
\usepackage{epsfig,rotating,setspace,latexsym,amsmath,epsf,amssymb,bm,theorem}
\usepackage{cite,psfrag,subfigure,algorithmic}
\usepackage{amsfonts}

\title{AWGN Channel under Time-Varying Amplitude Constraints with Causal Information at the Transmitter\thanks{This work was supported by NSF Grants CCF 07-29127, CNS 09-64632, CCF 09-64645, CCF 10-18185, and CNS 11-47811.}}

\author{Omur Ozel \qquad Sennur Ulukus \\
\normalsize Department of Electrical and Computer Engineering\\
\normalsize University of Maryland, College Park, MD 20742 \\
\normalsize {\it omur@umd.edu} \qquad {\it ulukus@umd.edu}}

\newtheorem{theorem}{Theorem}

\IEEEoverridecommandlockouts
\begin{document}

\maketitle

\begin{abstract}
We consider the classical AWGN channel where the channel input is constrained to an amplitude constraint that stochastically varies at each channel use, independent of the message. This is an abstraction of an energy harvesting transmitter where the code symbol energy at each channel use is determined by an exogenous energy arrival process and there is no battery for energy storage. At each channel use, an independent realization of the amplitude constraint process is observed by the transmitter causally. This scenario is a state-dependent channel with perfect causal state information at the transmitter. We derive the capacity of this channel using Shannon's coding scheme with causal state information. We prove that the code symbols must be selected from a finite set in the capacity achieving scheme, as in the case of Smith. We numerically study the binary on-off energy arrivals where the amplitude constraint is either zero or a non-zero constant.
\end{abstract}

\section{Introduction}
We consider a communication scenario where transmission energy is not available at the transmitter before the communication starts; instead, it arrives from an exogenous energy source during the course of communication as an independent and identically distributed (i.i.d.) process. Moreover, the transmitter does not have a battery to store the arriving energy. Hence, the code symbol energy in a channel use is constrained to the energy arrived in that channel use. We assume that the arrived energy can be observed by the transmitter causally, right before the code symbol is decided. Therefore, the code sequence is a function of the observed energy arrival and we find the capacity of this time-varying amplitude constrained system with causal state information at the transmitter.

\begin{figure}[t]
\begin{center}
\includegraphics[width=0.97\linewidth]{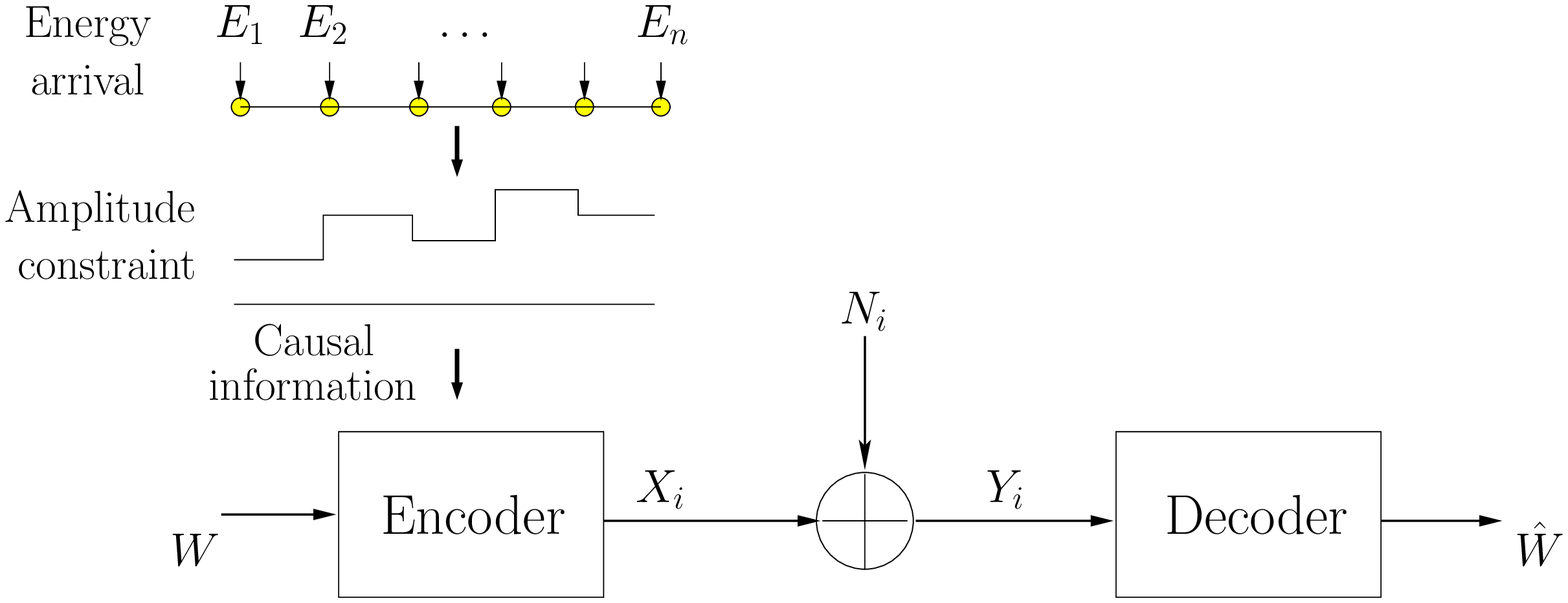}
\end{center}
\caption{The AWGN channel with time-varying amplitude constraints.}
\label{sys1}
\end{figure}

We consider the classical scalar AWGN channel model with the input-output relationship 
\begin{align} 
Y_i = X_i + N_i, \qquad i=1,\ldots,n 
\end{align} 
where $X_i$ is the channel input, $N_i$ is the i.i.d. Gaussian noise with zero-mean and unit-variance and $Y_i$ is the channel output at the $i$th channel use. $E_1,\ldots,E_n$ is the i.i.d. energy arrival sequence where $E_i \in \mathcal{E}$ and $|\mathcal{E}|$ is finite. The system model is shown in Fig. \ref{sys1}. At each channel use, the transmitter observes $E_i$ and generates a channel input $X_i$ that satisfies $X_i^2 \leq E_i$, i.e., the code symbol is amplitude constrained to (the square root of) the observed energy. Therefore, the major effect of energy arrivals is the time variation in the amplitude constraint that the code symbol should obey at each channel use. As the transmitter can observe the energy arrival causally, the resulting system is a state-dependent channel with causal state information at the transmitter and no state information at the receiver. The state of the channel is the amount of energy available at each channel use. At each state, the channel conditioned on the realized state is an AWGN channel with an input amplitude constraint equal to the square root of the arrived energy. 


The channel capacity of the static amplitude constrained AWGN channel was first studied by Smith in \cite{Smith71}. In \cite{Smith71}, Smith proved that the capacity achieving input distribution has a finite support set and presented an algorithm for the calculation of that distribution. This line of research has later been extended in \cite{Shamai95,abufaycal,gursoy, meyn} for various channels including quadrature-amplitude constrained AWGN channel and Rayleigh and Ricean fading channels. In \cite{kschischang}, the finiteness of the support set of the capacity achieving distribution for conditionally Gaussian channels with bounded inputs, which encompasses a large class of practical channels, is proved. In particular, optical channels and fading MIMO channels with and without state information at the receiver are encompassed in the finiteness result of \cite{kschischang}. Moreover, \cite{singh09} reports finiteness of the capacity achieving distribution for the quantized output AWGN channel. Also in \cite{guo}, capacity achieving input distribution for a duty cycle constrained system is shown to have countably infinite mass points with finite number of points in each bounded interval. In \cite{aslan04}, new sufficient conditions for the noise distribution are provided for the optimality of discrete channel inputs in an amplitude constrained additive noise channel. 

The capacity and optimal coding for a state-dependent channel with causal state information at the transmitter and no state information at the receiver was characterized by Shannon in \cite{Shannon58IBM}. In \cite{Shannon58IBM}, Shannon proved that the capacity of the state-dependent channel with causal state information available at the transmitter only is equal to the capacity of an equivalent channel which has an input alphabet extended by the cardinality of the state alphabet. In the capacity achieving coding scheme, the codewords are matrices rather than vectors, whose number of columns is the block length and the number of rows is the cardinality of the state alphabet. At each channel use, the code symbol that corresponds to the observed state is put to the channel. In the sequel, we refer to this coding scheme as the {\it Shannon strategy}.

The problem that we wish to address in this paper has two main characteristics:
\begin{enumerate}
\item amplitude constraints due to available energy, and
\item a state-dependent channel due to different energy arrivals where the state is naturally known to the transmitter but not to the receiver.
\end{enumerate}
In this paper, we find the capacity achieving scheme for the AWGN channel with time-varying amplitude constraints by applying and extending the results of Smith \cite{Smith71} and Shannon \cite{Shannon58IBM}. We obtain the capacity by applying the Shannon strategy to the time-varying amplitude constrained channel and optimizing the input distribution of the resulting extended alphabet channel. In particular, we extend the alphabet of the channel in accordance with the amplitude constraints and construct an equivalent channel which has the number of inputs equal to the cardinality of the alphabet of energy arrivals. Each input variable is constrained in amplitude by the square root of the corresponding amount of energy. We show that in the capacity achieving distribution each input has a finite support set, extending the result of Smith \cite{Smith71}. 

Next, we numerically study the considered setting with a specific binary energy arrival process, which we refer to as {\it on-off} energy arrivals: at each channel use, either $E$ amount of energy arrives or no energy arrives. We compare the capacity with several upper bounds such as the capacity of the AWGN channel with state information available at both sides and the capacity of the AWGN channel with an unlimited energy buffer (i.e, an infinite sized battery) \cite{ozel_pimrc,it-submit}. The numerical results show that the capacity with an infinite sized battery is considerably higher than the capacity with no battery (i.e., the setting of this paper), indicating the usefulness of collecting energy. In \cite{tobesubmitted}, we extend the framework considered here from i.i.d. to general stationary and ergodic energy arrival processes, and to additive noise channels with the general family of noise distributions considered in \cite{aslan04}.

\section{Time-Varying Amplitude Constraints}
\label{tv}
Before we address the time-varying amplitude constraints, we start with providing the necessary preliminary background on the AWGN channel with a static amplitude constraint.  

\subsection{Static Amplitude Constrained AWGN Channel}
\label{ss}
Discreteness of the capacity achieving input distribution for a static amplitude constrained AWGN channel was first shown by Smith in \cite{Smith71} and later the result was generalized for a wide class of noise distributions in \cite{aslan04}. The channel capacity under the amplitude constraint $A$ is \cite{Smith71} \begin{align} C_{Sm}(A) = \max_{F \in \mathcal{F}} I_{F}(X;Y) \end{align} where $\mathcal{F}$ is the space of input probability distribution functions whose support sets are constrained to $[-A,A]$. The subscript {\it Sm} refers to Smith. The proof of finiteness of the distribution is by contradiction and it requires systematic application of several steps \cite{Smith71,aslan04}: The first step is to prove that the space of distribution functions is compact and the objective function is concave in the weak topology. Next step is to obtain an optimality condition in terms of the support set. Smith, then, shows that the mutual information density function is analytic in a subset of $\mathbb{C}$ that includes $\mathbb{R}$. Finally, he shows that the optimality condition, after using identity theorem from complex analysis, leads to a contradiction.

\subsection{Time-Varying Amplitude Constrained AWGN Channel}

We now consider the time-varying amplitude constrained channel shown in Fig. \ref{sys1}. Let $A$ be the amplitude constraint random variable with the alphabet $\mathcal{A}=\{a_1,\ldots,a_M\}$. $\{A_k\}_{k=1}^\infty \in \mathcal{A}$ is the i.i.d. amplitude constraint process with probability that $A_k = a_i$ is equal to $p_{i}$ for all $k$. The realizations of the amplitude constraints $\{A_1,\ldots,A_n\}$ are observed by the transmitter causally and the code symbol must comply with the observed amplitude constraint at each channel use: 
\begin{align} 
|X_k| \leq A_k, \qquad k=1,\ldots,n
\end{align} 
The receiver has no information about the variations of the amplitude constraint at the transmitter. This is a state-dependent channel with causal state information available at the transmitter only \cite{Shannon58IBM}. The code sequence is determined as a function of the observed amplitude constraint sequence and the channel capacity is
\begin{align} 
\label{fr}
C_{Sh} = \max_{p_T(t)} I(T;Y)
\end{align} 
where $T=[T_1,\ldots,T_M]$ is an extended channel input related with the output as
\begin{align}
p_{Y|T}(y|t)=\sum_{i=1}^M p_i p_N(y-t_i)
\end{align}
where $p_N(.)$ is the Gaussian density. In (\ref{fr}), the subscript {\it Sh} refers to Shannon. 
\subsection{Other Scenarios with Side Information}

We can cast the arriving energy as a state variable attached to the channel. Knowledge of the state affects the achievable rates. If neither the transmitter nor the receiver has perfect knowledge of $\{A_k\}$, coding should be performed according to the minimum of the possible amplitude constraints otherwise some code symbol will violate amplitude constraint at some channel use. Hence, the best achievable rate in this case is 
\begin{align} 
C_{no-si} = C_{Sm}(\min_i a_i) 
\end{align}
where $C_{Sm}(\min_i a_i)$ is Smith's capacity under the amplitude constraint $\min_i a_i$.

On the other extreme, if perfect information of amplitude constraints is available at both the transmitter and the receiver, a multiplexed coding strategy achieves the maximum possible rate. Consider $|\mathcal{A}|$ parallel codes each generated with code rates $R_i$ subject to the amplitude constraint $a_i$. Since the receiver can track the changes in the amplitude constraint perfectly and due to the inherent stationarity of the amplitude constraint process, multiplexing codes results in the following achievable rate: $\sum_i p_i R_i = E_{A}[R(A)]$. Maximizing over each amplitude constraint, we obtain the capacity in this case as 
\begin{align} 
C_{si@both}=\sum_{i=1}^{|\mathcal{A}|}p_iC_{Sm}(a_i) 
\end{align} 
where $C_{Sm}(a_i)$ is Smith's capacity under the amplitude constraint $a_i$.

\section{Capacity with Causal Information at the Transmitter}

For simplicity, we will assume that the amplitude constraint process takes two different values, $a_1$ and $a_2$ with probabilities $p_{1}$ and $p_{2}=1-p_{1}$. However, the analysis is valid for any finite value of $|\mathcal{A}|$. 

Before dealing with the capacity achieving input distribution, we describe the capacity achieving scheme \cite{Shannon58IBM}. Codewords are matrices of $n$ columns (block length) and of two rows (number of states). The columns are generated as $n$ i.i.d. realizations of $T=(T_1,T_2)$ with the optimal cdf $F^*$. Then, $2^{nR}$ such code matrices are generated as the codebook and they are announced to the receiver. During the message transmission, as the causal information of the amplitude constraint is observed, the transmitter puts the code symbol at the corresponding row to the channel. Decoding is performed using joint typicality, and as $n \rightarrow \infty$, the probability of error tends to zero if $R<C$.

We will now find the optimal cdf $F^*$. We define the random variable $T$ over $\mathcal{T}=[-a_1,a_1]\times[-a_2,a_2]$ as $T = [T_1,T_2]$ where $T_1$ and $T_2$ have support sets $[-a_1,a_1]$ and $[-a_2,a_2]$ with joint cumulative distribution function $F$. The space of joint probability distribution functions over $[-a_1,a_1] \times [-a_2,a_2]$ is: \begin{align} \Omega \triangleq \left\{ F : \int_{-a_1}^{a_1}\int_{-a_2}^{a_2} dF(t_1,t_2) = 1\right\} \end{align} 
The capacity of the AWGN channel with time-varying amplitude constraint and causal information at the transmitter is: 
\begin{align} C = \max_{F \in \Omega} I_F(T;Y) \end{align} with \begin{align} I_F(T;Y)= \int_{-a_1}^{a_1}\int_{-a_2}^{a_2} \int_{-\infty}^{\infty} f(y|t_1,t_2)\log\left(\frac{f(y|t_1,t_2)}{f(y;F)}\right) dydF\end{align} 
where \begin{align} f(y|t_1,t_2)&=p_{1}p_{N}(y-t_1)+p_{2}p_N(y-t_2) \\ f(y;F)&=\int_{-a_1}^{a_1}\int_{-a_2}^{a_2} f(y|t_1,t_2)dF(t_1,t_2) \end{align}
Plugging in the AWGN expressions for the original channel, we will solve the following optimization problem: \begin{align}\label{opt_prob} \max_{F\in \Omega} \int_{-a_2}^{a_2} &\int_{-a_1}^{a_1}\int_{-\infty}^{\infty} \log\left(\frac{\sum_{i=1}^2\frac{p_i}{\sqrt{2\pi}}e^{-\frac{(y-t_i)^2}{2}}}{f(y;F)}\right)\nonumber \\ & \left(\sum_{i=1}^2\frac{p_i}{\sqrt{2\pi}}e^{-\frac{(y-t_i)^2}{2}}\right)dydF(t_1,t_2)  \end{align}
    
We will establish that the capacity achieving input distribution has a support set of finite cardinality. We will apply the steps of \cite{Smith71, aslan04} for the proof, which are summarized in Section \ref{ss}. The main difference between the static amplitude constrained and the time-varying amplitude constrained problems resides in the fact that the channel between $T$ and $Y$ is not an additive channel. Hence  $f(y;F)$ is not obtained through a convolution integral and $h(Y|T=(t_1,t_2))$ is not a constant, it takes different values at different $(t_1,t_2)$. However, we will show that it is possible to obtain similar arguments for the line $t_1=t_2$ and this will lead us to the claimed extension. This way, we also extend the class of problems for which this technique is applicable.

We note that $I_F(T;Y)$ is a concave functional of $F\in \Omega$. Moreover, $\Omega$ is a convex and compact space in the weak topology. Finally, $I_F(T;Y)$ is strictly concave and weakly differentiable in $\Omega$ with the mutual information density $i(t_1,t_2;F)$ such that the derivative at $G \in \Omega$ is 
\begin{align} 
\frac{d}{dF}I_{G}(T;Y)=&\int_{-a_2}^{a_2} \int_{-a_1}^{a_1} i(t_1,t_2;F)dG(t_1,t_2) - I_F(T;Y)
\end{align} 
In particular, the mutual information density is \begin{align} \label{mutdens} i(t_1,t_2;F)=\int_{-\infty}^\infty\log\left(\frac{f(y|t_1,t_2)}{f(y;F)}\right)f(y|t_1,t_2)dy \end{align}
Moreover, $i(t_1,t_2;F)$ in (\ref{mutdens}) is continuous and has analytic extension over the two-dimensional complex numbers $\mathbb{C}^2$. These claims are parallel to those in \cite{Smith71,aslan04,kschischang}. The proofs of these and all upcoming claims can be found in \cite{tobesubmitted}. The convexity and compactness of $\Omega$ as well as the concavity and weak differentiability of $I_F(T;Y)$ guarantee the uniqueness of the solution of the optimization problem in (\ref{opt_prob}). Next, we have the following Lagrangian theorem:
\begin{theorem}
\label{necsuf}
Let $F^* \in \Omega$ and let $\mathcal{S}_{F^*}$ indicate the support set of $F^*$. Then, $F^*$ is optimal if and only if \begin{align} i(t_1,t_2;F^*) &\leq C, \ \forall (t_1,t_2) \in [-a_1,a_1] \times [-a_2,a_2] \\ i(t_1,t_2;F^*) &= C, \ \forall (t_1,t_2) \in \mathcal{S}_{F^*} \end{align} where $C=I_{F^*}(T;Y)$. 
\end{theorem}    

We now prove that $\mathcal{S}_{F^*}$ is finite by contradiction. Assume $S_{F^*}$ includes infinitely many elements in $\mathbb{R}^2$. By using a sequence of steps analogous to those in \cite{Smith71,aslan04,kschischang}, we obtain the following result:
\begin{align}
i(z_1,z_2) = C, \qquad (z_1,z_2) \in \mathbb{C}^2 
\end{align}
where $C=I_{F^*}(T;Y)$ is the optimal value for the problem in (\ref{opt_prob}), i.e., the capacity. In particular, $i(t_1,t_2) = C$ for all $(t_1,t_2) \in \mathbb{R}^2$. For $t_1=t_2=t$, we obtain $\forall t \in \mathbb{R}$
\begin{align} \label{this}
\int_{-\infty}^{\infty} \frac{1}{2}e^{-\frac{(y-t)^2}{2}}\log\left(p(y;F)\right) dy = -C - \frac{1}{2}\log(2\pi e)
\end{align}
(\ref{this}) causes a contradiction as in \cite{Smith71,aslan04,kschischang}. Therefore, we have the following theorem. 
\begin{theorem}\label{fnt}
$\mathcal{S}_{F^*}$ is a finite set.
\end{theorem}

\subsection{Algorithm to Find $\mathcal{S}_{F^*}$}
\label{ssec}
We use a procedure similar to that in \cite{Smith71,kschischang} to find the finite support set of the capacity achieving distribution. We first fix the cardinality of $\mathcal{S}_{F}$. The problem becomes a $2(|\mathcal{S_{F}}|)-1$ variable convex optimization problem with convex constraints. After solving this optimization problem, we get the distribution $F$ that optimizes the objective function subject to the fixed cardinality. If $F$ satisfies the necessary and sufficient conditions in Theorem \ref{necsuf}, then $F^*=F$. Otherwise, we increase the cardinality $|\mathcal{S}_F|$ by one and repeat the procedure. We start with assuming that $|\mathcal{S}_{F}|=2$. The optimal distribution subject to this cardinality constraint has two mass points at the edges $(t_1,t_2)= (a_1,a_2)$ and $(t_1,t_2) = (-a_1,-a_2)$. If this input distribution does not satisfy the necessary and sufficient conditions in Theorem \ref{necsuf}, we solve the problem subject to $|\mathcal{S}_{F}|=3$. We repeat the procedure until we reach a distribution that satisfies the conditions in Theorem \ref{necsuf}.

\section{On-Off Energy Arrivals}

In this section, we consider on-off energy arrivals: At each channel use either $E$ units of energy arrives or zero energy arrives. In this case, $\mathcal{A}=\{a_1,a_2\}$ where $a_1=0$ and $a_2=\sqrt{E}>0$. We denote the probability that $E$ units of energy is harvested with $p_{on}$. We have \begin{align} \label{yr} C_{si@both}=p_{on}C_{Sm}(\sqrt{E}) \end{align} Note that $C \leq C_{si@both}$ and $C_{no-si}=0$. 

$\mathcal{S}_{F^*} \subset \mathbb{R}^2$ is such that for all $(t_1,t_2) \in \mathcal{S}_{F^*}$, $t_1=0$ since the amplitude constraint for $T_1$ is $a_1=0$. Hence, the optimization problem is over the one dimensional cdf $F_{T_2}(t_2)$ only. The space of cdfs $\mathcal{F}_{T_2}$ is such that \begin{align} \int_{-\sqrt{E}}^{\sqrt{E}} dF_{T_2}(t_2)=1 \end{align} For $p_N(y)=\frac{1}{\sqrt{2\pi}}e^{-\frac{y^2}{2}}$, we have the capacity  
\begin{align} C = \max_{F_{T_2} \in \mathcal{F}_{T_2}} I_{F_{T_2}}(T_2;Y) \end{align} 
with 
\begin{align} 
I_{F_{T_2}}(T_2;Y)= \int_{-\sqrt{E}}^{\sqrt{E}} \int_{-\infty}^{\infty} f(y|t_2)\log\left(\frac{f(y|t_2)}{f(y;F_{T_2})}\right) dydF_{T_2}
\end{align} 
where 
\begin{align}\label{fyx} 
f(y|t_2)&=(1-p_{on})p_{N}(y)+p_{on}p_N(y-t_2) 
\end{align}

Note that similar to the static amplitude constrained AWGN channel \cite{Smith71}, if $\sqrt{E}$ is small, the support set of $F_{T_2}^*$ is symmetric binary with two mass points located at $\pm \sqrt{E}$. For $p_{on}=1$, the problem reduces to Smith's amplitude constrained AWGN capacity problem. In this case, using the algorithm in Section \ref{ssec}, we find that if $\sqrt{E}\leq 1.66$, then symmetric binary distribution for $T_2$ is optimal and if $\sqrt{E}>1.66$, optimal distribution of $T_2$ has more than two mass points. For $p_{on}\neq 1$, the channel between $T$ and $Y$ is different from an AWGN channel; hence, the optimizing distribution is different. To capture this effect, we define a function of $p_{on}$ as follows \begin{align} \nonumber U(p_{on})=\max\{x\in \mathbb{R}: g(t_2,x) \leq g(x,x), \ \forall t_2 \in [-x,x]\} \end{align} where $g(t_2,x)$ is the mutual information density $i(x,t_2;F)$ evaluated at the binary symmetric distribution with two equiprobable mass points located at $t_2=-x$ and $t_2=x$. Here, $f(y|t_2)$ is given by (\ref{fyx}). In view of the conditions in Theorem \ref{necsuf}, $U(p_{on})$ is the highest amplitude constraint under which the binary symmetric distribution is optimum when the energy arrival probability is $p_{on}$. 

\begin{figure}[t]
\begin{center}
\includegraphics[width=0.93\linewidth]{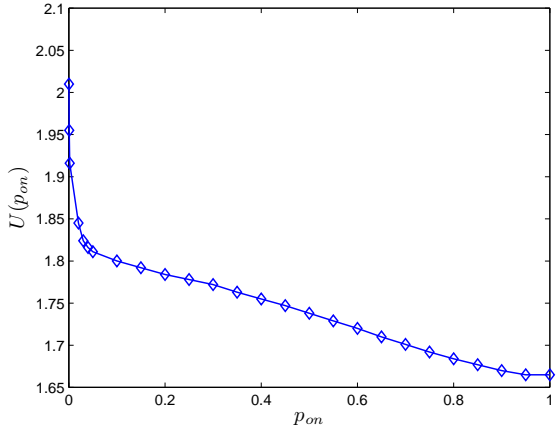}
\end{center}
\caption{$U(p_{on})$ function for the AWGN channel with unit noise power.}
\label{fig1}
\end{figure}

The function $U(p_{on})$ is monotonically decreasing with $p_{on}$ as shown in Fig. \ref{fig1}. As $p_{on}$ is decreased, the number of channel uses the nature allows the transmitter to send a non-zero data symbol decreases and this leads to smaller capacity. We observe in Fig. \ref{fig1} that as $p_{on}$ is increased, binary input distribution becomes optimal for a smaller range of amplitude constraints, leading $U(p_{on})$ to be monotonically decreasing.

\begin{figure}[t]
\begin{center}
\includegraphics[width=0.93\linewidth]{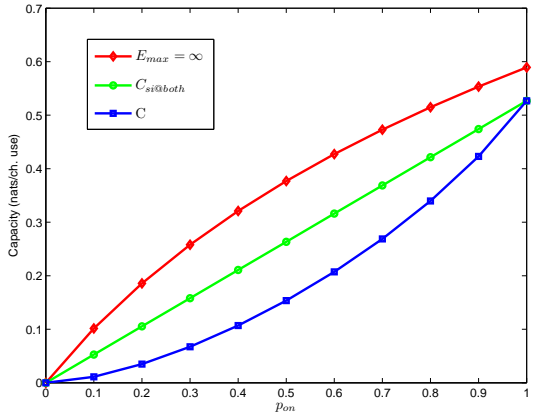}
\end{center}
\caption{Capacity versus $p_{on}$ for $E=2.25$, i.e., $\sqrt{E}=1.5$.}
\label{fig2}
\end{figure}

\begin{figure}[t]
\begin{center}
\includegraphics[width=0.98\linewidth]{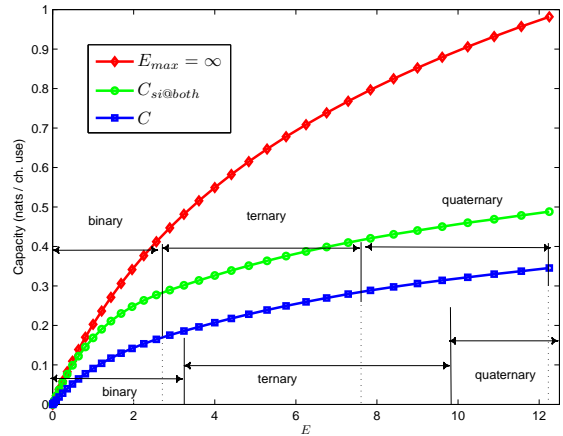}
\end{center}
\caption{Capacity versus $E$ when $p_{on}$=0.5.}
\label{fig3}
\end{figure}

In Figs. \ref{fig2} and \ref{fig3}, we compare the channel capacity under the on-off energy arrival when the state information is available at the transmitter causally with the capacity when the state (energy arrival) information is available at both sides. We also plot the channel capacity when the battery size is unlimited, i.e., $E_{max}=\infty$, as in \cite{ozel_pimrc,it-submit}. The capacity in this case is $\frac{1}{2}\log\left(1+p_{on}E\right)$. In Fig. \ref{fig2}, we observe the differences in the capacities for different values of $p_{on}$ when $\sqrt{E}=1.5$. The capacity achieving input distribution is binary for all $p_{on}$ in this case since $\sqrt{E}=1.5 < 1.66$. In Fig. \ref{fig3}, we plot the capacities for different $E$ for a fixed $p_{on}$. Note that the capacity achieving input distribution changes as $E$ is increased. We show the ranges over which the capacity achieving input distribution is binary, ternary and quaternary in Fig. \ref{fig3}. In particular, the capacity achieving distribution for (\ref{yr}) is the capacity achieving distribution with a constant amplitude constraint $\sqrt{E}$. We observe that the transition from binary to ternary for $C_{si@both}$ occurs at $E=(1.66)^2$ while it occurs for the capacity $C$ with causal state information at the transmitter at $E=(U(p_{on})|_{p_{on}=0.5})^2 = (1.74)^2$. We also observe that as $E$ gets large, the capacity with an unlimited battery is significantly larger than the capacities with no battery, with or without the state information at the receiver.  

\section{Conclusion}
In this paper, we considered the capacity of the AWGN channel with time-varying amplitude constraints. This scenario represents an energy harvesting communication system which uses harvested energy for data transmission and which has no battery to store the energy for future use. The energy arrivals impose amplitude constraints on the code symbol at each time. We found the capacity of this system.  We showed that the capacity achieving coding is performed by choosing code symbols from a finite set among the allowable continuum of points at each energy state. Next, we studied the {\it on-off} energy arrivals where the amplitude constraint is either zero or a non-zero value. We provided numerical illustrations for the capacity of the on-off energy arrivals. The numerical results indicate the capacity gain provided by an unlimited battery.

\end{document}